\documentclass[journal]{IEEEtran}

\ifCLASSINFOpdf
\else
   \usepackage[dvips]{graphicx}
\fi
\usepackage{url}

\usepackage{graphicx}
\usepackage{amsmath,amssymb,amsfonts}

\usepackage[linesnumbered,boxed,ruled,commentsnumbered]{algorithm2e}

\usepackage{cite}

\usepackage{epstopdf}

\usepackage{bm}
\usepackage{booktabs}

\usepackage{subfigure}
\usepackage{enumerate}
\usepackage{makecell}

\makeatletter
\renewcommand{\maketag@@@}[1]{\hbox{\m@th\normalsize\normalfont#1}}
\makeatother

\usepackage{xcolor}

\usepackage{ragged2e}

\begin{document}

\title{Symplectic Optimization for Cross Subcarrier Precoder Design with 
	Channel Smoothing in Massive MIMO-OFDM System
}

\author{Yuxuan Zhang, \IEEEmembership{Student Member, IEEE,} An-An Lu, \IEEEmembership{Member, IEEE,} and 
	Xiqi Gao, \IEEEmembership{Fellow, IEEE}
	
	\thanks{Y. X. Zhang, A.-A. Lu and X. Q. Gao are with the National Mobile Communications Research Laboratory, Southeast University,
		Nanjing, 210096 China, 
		and also with Purple Mountain Laboratories, Nanjing 211111, China, 
		e-mail: y\_x\_zhang@seu.edu.cn, aalu@seu.edu.cn,  xqgao@seu.edu.cn.}
}

\maketitle

\begin{abstract}
	In this paper, we propose a cross subcarrier precoder design (CSPD) for massive multiple-input multiple-output (MIMO) orthogonal frequency division multiplexing (OFDM) systems. The aim is to maximize the weighted sum-rate (WSR) performance while considering the smoothness of the frequency domain effective channel. To quantify the smoothness of the effective channel, we introduce a delay indicator function to measure the large delay components of the effective channel. An optimization problem is then formulated to balance the WSR performance and the  delay indicator function. By appropriately selecting the weight factors in the objective function and  the parameters in the  delay indicator function, the delay spread of the effective channel can be reduced, thereby enhancing the smoothness of the effective channel. To solve the optimization problem, we apply the symplectic optimization, which achieves faster convergence compared to the gradient descent methods. Simulation results indicate that the proposed algorithm achieves satisfying WSR performance while maintaining the smoothness of the effective channel.
\end{abstract}

\begin{IEEEkeywords}
	Massive MIMO, precoder design, channel smoothness, symplectic optimization
\end{IEEEkeywords}

\IEEEpeerreviewmaketitle

\section{Introduction}
Massive multiple-input multiple-output (MIMO) combined with orthogonal frequency division multiplexing (OFDM) is a foundational technology for the fifth-generation (5G) cellular communication systems. It is anticipated to have a significantly enhanced impact in future sixth-generation (6G) networks \cite{8804165,8861014,9003618}. Massive MIMO-OFDM systems, leveraging large antenna arrays at the base station (BS) to simultaneously serve multiple users, achieve significant enhancements in spectral and energy efficiency. Extensive research has been conducted on key aspects of these systems, including channel estimation, precoding, and user positioning \cite{10026502,9364875,8753608}. One of the major challenges in such systems is the management of the multi-user interference, which can severely degrade the system performance. Hence, efficient precoder design at the BS is essential for optimizing the system throughput \cite{luo,ge,zhang,zhang_sum-rate-optimal_2022}.

In massive MIMO systems, nonlinear precoders like dirty-paper coding (DPC) \cite{1683918} can achieve optimal performance but suffer from excessive computational complexity. Alternatively, linear precoders, including maximum ratio transmission (MRT) \cite{6812124}, regularized zero-forcing (RZF) precoders \cite{8668481}, minimum mean-squared error (MMSE) precoders \cite{1046557}, and signal-to-leakage-and-noise ratio (SLNR) precoders \cite{zhang_sum-rate_2017}, are widely used for their favorable trade-off between complexity and performance.
Meanwhile, the weighted MMSE (WMMSE) precoders \cite{christensen_weighted_2008} that maximizes the weighted sum-rate (WSR)  has also been extensively studied.
When combining OFDM with massive MIMO, these precoders are applied to each subcarrier independently. 
However, the per-subcarrier precoder designs disrupt the smoothness of the effective channel (formed by the concatenation of the precoder and the channel) across  adjacent subcarriers, which adversely affects channel estimation performance at the receiver \cite{smoothexp1}.

To maintain the smoothness of the effective channels in OFDM systems, a common method is to use the same precoder for all subcarriers \cite{hu_wideband_2021}, but its performance is degraded due to the frequency-selectivity of the channel. Subcarrier grouping methods, which use the same precoder for small groups of adjacent subcarriers \cite{venugopal_optimal_2019}, exhibit better performance. However, the discontinuities at the edges of subcarrier groups are introduced, and cause performance degradation of the effective channel estimation. To address this issue, a joint transceiver design that generates a smooth beam-steering matrix for wireless local area network (WLAN)  is  proposed  \cite{9348441}.   Furthermore, phase-rotated SVD \cite{smooth} has been employed to improve effective channel estimation.
To maximize the  sum-rate while maintain the smoothness of the effective channels, the transform domain precoding vectors are designed in \cite{9348} and converted to the frequency domain for different subcarriers. 
The symplectic optimization \cite{symplectic2}, which draws from concepts in Hamiltonian dynamics and symplectic integration, is then applied to solve an unconstrained sub-problem to reduce complexity. 
However, the effectiveness of using transform domain precoding vector is not justified rigorously.

In this paper, we focus on the  precoder design  that jointly optimize  the WSR performance  and the  smoothness of the effective channel.
We optimize the original precoder directly.
To measure the smoothness of the effective channel, we define a delay indicator function that characterizes the energy associated with the large delay components of the effective channel in the delay domain. Further, we formulate an  optimization problem with weighted sum of two functions to balance the WSR performance and the  delay indicator function. By setting reasonable weight factor in the objective function and the parameter in the  delay indicator function, the delay spread can be reduced and the effective channel can be smooth. Then, we apply the optimization via conformal Hamiltonian systems \cite{jordandan}, which can also be viewed as a kind of symplectic optimization,  to solve the constrained optimization problem and provide an  algorithm. The proposed algorithm achieves a faster rate than gradient descent. Simulation results indicate that the proposed precoder achieves a satisfying WSR performance while maintaining the smoothness of the effective channel.  Meanwhile, the proposed precoder design exhibits lower computational complexity in comparison with traditional precoder designs.

The rest of this paper is orgnized as follows:  Section II presents the  problem
formulation of CSPD. Section III proposes the precoder design with symplectic optimization. Section IV presents simulation results illustrating the benefits
of the CSPD. The paper concludes in Section V.

\emph{Notations}: 
The  conjugate, transpose and hermitian of matrix $\mathbf A$ are denoted as $\mathbf{A}^*$, $\mathbf{A}^T$ and $\mathbf{A}^H$, respectively.   The trace of $\mathbf{A}$ is represented as the operator  ${\rm{tr}}(\mathbf{A})$.  
The identity matrix of size $M \times M$ is denoted as $\mathbf{I}_M$.  The operation $\rm{diag} (\mathbf{a})$ constructs a square diagonal matrix with the elements of vector $\mathbf{a}$ on its main diagonal. Additionally, $\mathbf{A} = \rm{Bdiag}\{\mathbf{A}_1, \cdots, \mathbf{A}_N\}$ creates a block diagonal matrix with blocks $\mathbf{A}_1, \cdots, \mathbf{A}_N$ along its main diagonal.

\section{Problem Formulation}

In this section, we formulate the CSPD as an optimization problem which balances the WSR performance and the smoothness of the  effective channels.

\subsection{System Model}
We consider a single-cell massive multiple-input multiple-output (MIMO) system consists of a BS equipped with a uniform planar array (UPA) comprising $M=M_zM_x$ antennas, where $M_x$ and $M_z$ denote the number of horizontal and vertical antennas, respectively. The BS serves 
$K$ single-antenna users. Orthogonal frequency division multiplexing (OFDM) modulation is employed. The number of subcarriers is
$N_c$, and
$N_v$ subcarriers are allocated for data transmission. Each time slot contains $N_b$ OFDM symbols, and the channel parameters remain constant within each OFDM symbol but vary across different symbols. 
The diagram of the massive MIMO system is depicted in Fig.1

\begin{figure}[t]
	
	\centering
	
	\includegraphics[width=85mm,height=45mm]{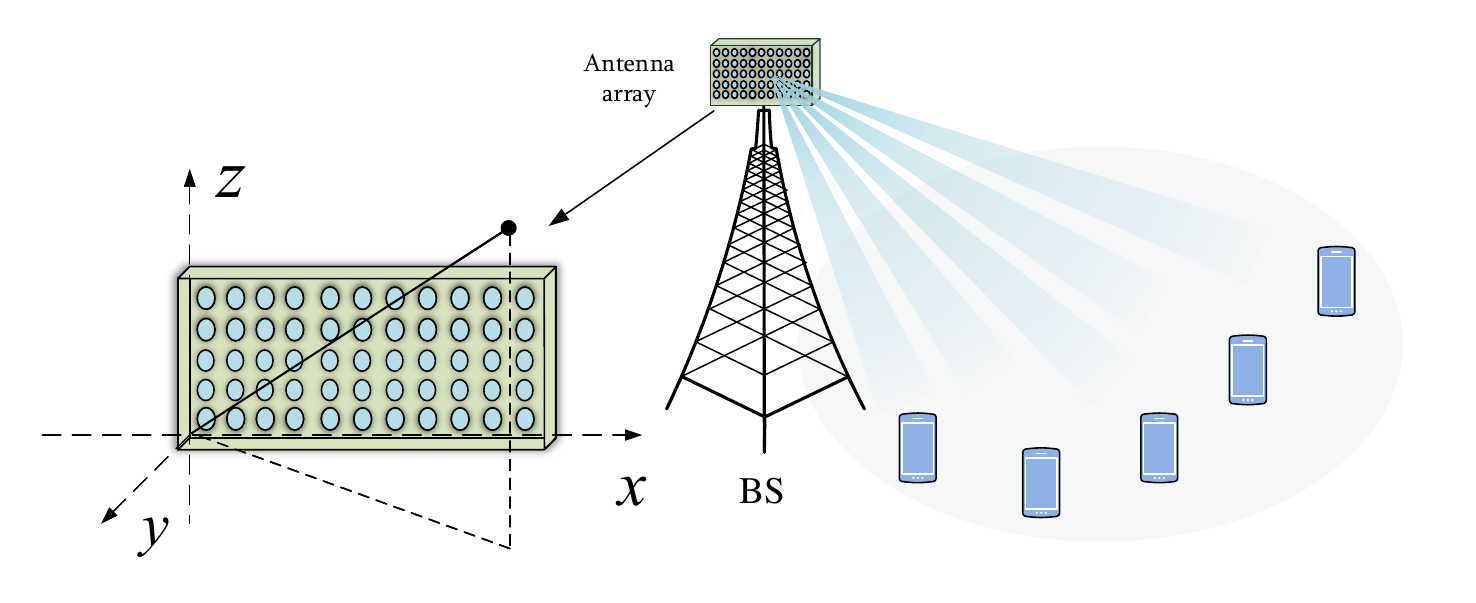}
	
	\caption{The massive MIMO system.}
	
	\label{Layout}
	
\end{figure}

\subsection{ Per-subcarrier Precoder Design}
Let $\mathbf p_{k}^c \in \mathbb{C}^{M \times 1}$ and $x_{k,c}$ denote  the precoding vector and the transmitted signal of the $k$-th user at the $c$-th subcarrier, respectively. The transmitted signal after precoding can be presented as
\begin{align}
	\mathbf x_{k,c}= \sum\limits_{k=1}^K	 \mathbf p_{k}^c x_{k,c}. 
\end{align}
The channel vector from the BS to user $k$  at the $c$-th
subcarrier is defined as $\mathbf h_{k,c}$.  The received signal $y_{k,c}$ of user $k$ at the $c$-th subcarrier  can be presented as
\begin{align}
	y_{k,c}&=\mathbf h_{k,c}\mathbf x_{k,c}+ z_{k,c}\notag\\&
	=\mathbf h_{k,c} \mathbf p_{k}^c x_{k,c} +\mathbf h_{k,c} \sum\limits_{l \ne k}^K	 \mathbf p_{l}^c x_{l,c} +  z_{k,c}\label{permodel}
\end{align}
where  the  noise ${ z}_{k,c}$ is   distributed as $\mathcal{CN}(0, \sigma_z^2 )$.   We treat the aggregate interference-plus-noise $z_{k,c}^{'}=\mathbf h_{k,c} \sum\limits_{l \ne k}^K	 \mathbf p_{l}^c x_{l,c} +  z_{k,c}$ as Gaussian noise. Let $\Gamma_{k,c}$ be denoted as the variance of $z_{k,c}^{'}$, we have
\begin{align}
	\Gamma_{k,c}
	&=\sigma_z^2+\sum\limits_{l\neq k}^{K}\mathbf{h}_{k,c} \mathbf p_l^c(\mathbf p_l^c)^H \mathbf{h}_{k,c}^H. \label{r_kc_ori}
\end{align}
Assuming that user $k$ can obtain $\Gamma_{k,c}$. The  rate of user $k$ at the $c$-th subcarrier can be computed as
\begin{align}
	{\mathcal{R}_{k,c}} 
	=
	\mathrm{log}(1+{\Gamma}_{k,c}^{-1}\mathbf{h}_{k,c}\mathbf p_k^c(\mathbf p_k^c)^H \mathbf{h}_{k,c}^H)
	.
\end{align}
To maximize the WSR for each subcarrier, the problem formulation of the  conventional per-subcarrier precoder design with total power constraint is given as
\begin{align}
	\mathbf p_1^c,\cdots,\mathbf p_K^c&=\mathop {\arg \,\max } \,\sum\limits_{k = 1}^Kw_k {{\mathcal{R}_{k,c}}} \notag \\  
	&{\rm s.t.}\,\,\,\sum\limits_{k = 1}^{K}(\mathbf p_k^c)^H \mathbf p_k^c \le P_c \label{per_problem}
\end{align}
where $w_k$ is used to ensure the fairness of different users and $P_c$ is the  power budget at $c$-th subcarrier. 
Since the precoders in \eqref{per_problem} are designed independently for each subcarrier, the smoothness of the resulting frequency domain effective channels  is not satisfied. This implies a weak correlation among  effective channels, thereby complicating effective channel estimation and signal detection at the receiver.

\subsection{ CSPD with Channel Smoothing}

To significantly improve the performance of the channel estimation and signal detection, we optimize precoders across subcarriers, and the smoothness of the frequency domain effective channel is considered as a objective function in the optimization problem. The delay spread can be adjusted by setting related parameters.

Let $\mathbf p_k$ be a  vector stacked by $\mathbf p_k^c$  as $\mathbf p_k=[(\mathbf p_k^1)^T,\cdots,(\mathbf p_k^{N_v})^T]^T\in \mathbb{C}^{MN_v \times 1}$ and $\mathbf H_k$ be a block diagonal matrix defined as $\mathbf H_k={\rm Bdiag}\{\mathbf h_{k,1},\cdots,\mathbf h_{k,N_v}\}\in \mathbb{C}^{N_v \times MN_v}$.
By converting the frequency domain effective channel sequence along subcarriers to the delay domain, we obtain  the large delay components of  the delay domain effective channels as
\begin{align}
	\mathbf d_k=\mathbf E_{N_e}\mathbf M_{I}\mathbf H_k\mathbf p_k
\end{align}
where $\mathbf M_{I}$  is an IDFT matrix, and delay components exceeding $N_e$ are defined as large delay components.   The  matrix $\mathbf E_{N_e}$ is used to extract the large delay components. It is defined as a diagonal matrix whose first 
$N_e$ diagonal  entries are $0$, and the rest are $1$. 
Further, we define a delay indicator function as
\begin{align}
	f_h(\mathbf p_k^1,\cdots,\mathbf p_k^{N_v})={\rm{tr}}(\mathbf d_k\mathbf d_k^H).
\end{align}
As the  delay indicator function decreases, the large delay components diminish, making the frequency domain effective channel smoother. When the   delay indicator function becomes sufficiently small, 
the corresponding delay spread of the effective channel is not large than $N_e$.

Let $\mathbf p$ be a vector stacked by the vectors $\mathbf p_k$ as $\mathbf p=[(\mathbf p_1)^T,\cdots,(\mathbf p_K)^T]^T\in \mathbb{C}^{KMN_v \times 1}$.  Define functions $\phi$ and $d$  as the total power of all precoders and the  delay indicator function  for all users, respectively.  
The formulas for $\phi$ and $d$ are given by
\begin{align}
	\phi(\mathbf p)&= {\rm{tr}}(\mathbf p^H \mathbf p),\\
	d(\mathbf p)&=\sum\limits_{k = 1}^K	f_h(\mathbf p_k^1,\cdots,\mathbf p_k^{N_v}).\label{phi2}
\end{align}
Let the matrices $\mathbf H$ and $\mathbf M$ be defined as the block diagonal matrices $\mathbf H={\rm Bdiag}\{\mathbf H_1,\cdots,\mathbf H_K\}$ and $\mathbf M={\rm Bdiag}\{\mathbf E_{N_e}\mathbf M_I,\cdots,\mathbf E_{N_e}\mathbf M_I\}$, respectively. The function $d(\mathbf p)$ in \eqref{phi2} can be rewritten as $d(\mathbf p)= \mathbf p^H\mathbf H^H\mathbf M^H\mathbf M \mathbf H \mathbf p$.

We aim to maximize the WSR while smoothing the frequency domain effective channel under  the power constraint. To achive this goal, we  formulate an optimization problem with the objective being the weighted sum of two functions as
\begin{align}
	\mathop {\arg \,\min }\limits_{\mathbf p } \,f(\mathbf p)  + \alpha d(\mathbf p) \notag\\
	{\rm s.t.}\,\,\,\phi(\mathbf p) = P
	\label{problem_qos}
\end{align}
where a preset weight factor $\alpha$  is introduced to balance the two functions,   $-f(\mathbf p_1,\cdots,\mathbf p_K)=\sum_{k = 1}^K\sum_{c = 1}^{N_v}w_k {{\mathcal{R}_{k,c}}}$ is the  WSR and $P=\sum_{c = 1}^{N_v}P_c$ is the power budget.
If we set $\alpha=0$, the problem \eqref{problem_qos} reduces to jointly optimizing the precoders over all used subcarriers. 
By solving this optimization problem with large $\alpha$, $d(\mathbf p)$ can be small enough and the smoothness of the frequency domain effective channels can be improved. Moreover, we can modify the  parameter $N_e$ to adjust the delay spread of the effective channel.

\section{Precoder Design with Symplectic Optimization}

In this section, we first introduce a dissipative augmented Hamiltonian system related to the optimization problem \eqref{problem_qos}. Then, we use the symplectic integration to obtain a stable algorithm by discretizing the Hamiltonian system.

\subsection{Briefly Introduction of Symplectic Optimization}

Symplectic optimization connects optimization problems with disspative dynamics systems, where the objective function serves as the potential energy \cite{symplectic2}. 
The disspative dynamic system always converges to the minimum of the potential energy, and thus can be used to solve the optimization problem. 
Symplectic optimization 
preserves
the continuous symmetries of the original dynamical system, enabling larger step sizes. Consequently,
algorithms based on symplectic optimization can move more quickly.
This method achieves a provably faster rate than gradient descent and meets the oracle lower bound.
Additionally, symplectic optimization performs efficiently when applied to large-scale data analysis \cite{symplecticv1}. Since the optimization problem \eqref{problem_qos} involves large-scale vectors operations, the symplectic optimization is naturally suitable to solve this problem.

Let $g(\mathbf p)=f(\mathbf p)  + \alpha d(\mathbf p)$. We can reformulate the optimization problem \eqref{problem_qos} as
\begin{align}
	&\mathop {\arg \,\min }\limits_{\mathbf p\in\mathcal N } \,g(\mathbf p)   
	\label{manifold}
\end{align}
where the manifold $\mathcal N$ is defined as $\mathcal N=\{\mathbf p|\phi(\mathbf p) = P\}$. The motion of $\mathbf p$ is constrained by $\phi(\mathbf p)$.

To illustrate symplectic optimization, we will first introduce the augmented Lagrangian system  related to the optimization problem \eqref{manifold}. Since the augmented Lagrangian system is conservative, its dynamics  oscillate around the minimum of the  optimization problem. To generate dynamics that convergence to the minimum of the problem \eqref{manifold}, a dissipative augmented Hamiltonian system is then introduced. Finally, the symplectic integration over the space of a position coordinate and a momentum coordiante is introduced to obtain a stable algorithm by discretizing the dynamic system.

\subsection{Augmented Lagrangian Systems}

Suppose $\mathcal N$ is embedded in an Euclidean space. 
Let the precoder $\mathbf p$ be the position coordinates in the Lagrangian dynamic system, $T(\dot{\mathbf p})=\frac{1}{2}\dot{\mathbf p}^H\mathbf M \dot{\mathbf p}$ be the kinetic energy of the dynamic system and $g(\mathbf p)$ be its
potential energy, where $\mathbf M$ is defined as the mass matrix. The augmented Lagrangian is defined as
\begin{align}
	L(\mathbf p,\dot{\mathbf p})=T(\dot{\mathbf p})-g(\mathbf p)-\phi(\mathbf p)\lambda
\end{align}
where $\lambda$ is the Lagrange multiplier. The Euler-Lagrange equations
of the variational formulation problem for $\int_{0}^{t}L(\mathbf p,\dot{\mathbf p})dt $ are presented as \cite{symplectic3}
\begin{align}
	\frac{d}{dt}(\frac{\partial L}{\partial \dot{\mathbf p}})-\frac{\partial L}{\partial {\mathbf p}}=\mathbf 0.
\end{align}
It follows that
\begin{align}
	\mathbf M \ddot{\mathbf p}+\nabla g(\mathbf p)+G(\mathbf p)\lambda=\mathbf 0.
\end{align}
Then, a first order differential equation can be obtained as
\begin{align}
	\dot{\mathbf p}=\mathbf v \notag\\
	\mathbf M \dot{\mathbf v}=-\nabla g(\mathbf p)-G(\mathbf p)\lambda\notag\\
	\phi(\mathbf p)=0 \label{kktsys}
\end{align}
where $G(\mathbf p)=\frac{\partial \phi}{\partial \mathbf p}(\mathbf p^*)$ is the Jacobian matrix of constraint $\phi(\mathbf p)$ at point $\mathbf p$ and $\mathbf v$ can be seen as the velocity coordinate of the position $\mathbf p$.

Since the augmented Lagrangian is time independent, the resulted dynamics would oscillate around the desired minimum on the constrained manifold. To obtain dynamics which can rapidly converge to the desired minimum, we need to define dissipative Lagrangian systems, where the Lagrangian is time dependent.
Furthermore, the dynamic system is a continuous system. In practice, we have to discretize it to obtain an practical algorithm.  
However, discretizations of Lagrangian dynamics are often fragile \cite{symplectic2}. 
Fortunately, a stable discretization can be readily constructed  by exploiting the dual Hamiltonian form of the dynamics. In the following, we will introduce 
the dissipative augmented Hamiltonian system \cite{jordandan}.

\subsection{Dissipative Augmented Hamiltonian System}
In the Hamiltonian system,
the momentum coordinate is defined as
\begin{align}
	\mathbf q=\frac{\partial L}{\partial \dot{\mathbf p}}=\mathbf M \mathbf p
\end{align}
in place of the velocity coordinate $\mathbf v=\dot{\mathbf p}$. The Hamiltonian is defined as the total energy of the system, given as
\begin{align}
	H(\mathbf p, \mathbf q)=g(\mathbf p)+T(\mathbf q). \label{Hamilton}
\end{align}
The augmented Hamiltonian is then $H(\mathbf p, \mathbf q) + \lambda\phi(\mathbf p)$. Let $H_{\mathbf q}$ and $H_{\mathbf p}$ denote the partial
derivatives of the Hamiltonian with respect to $\mathbf q$ and $\mathbf p$. 
Accroding to the proof of Theorem VI.1.3 in \cite{symplectic3},
the system \eqref{kktsys} becomes 
\begin{align}
	&\dot{\mathbf p}= H_{\mathbf q}(\mathbf p, \mathbf q)\notag\\
	&\dot{\mathbf q}=- H_{{\mathbf p}}(\mathbf p, \mathbf q)-G(\mathbf p)^T\lambda\notag\\
	&\phi(\mathbf p)= P. \label{Hsys}
\end{align}
The Jacobian of $\phi$ is denoted as $G(\mathbf p)$.  By differentiating the constraint once and twice with respect to time $t$, we have
\begin{align}
	&G(\mathbf p) H_{{\mathbf q}}(\mathbf p, \mathbf q)=0  \\
	&\frac{\partial}{\partial \mathbf p}\Big(G(\mathbf p) H_{{\mathbf q}}(\mathbf p, \mathbf q)\Big)H_{\mathbf q}(\mathbf p, \mathbf q)\notag\\&-G(\mathbf p)\frac{\partial}{\partial\mathbf q}\Big( H_{\mathbf q}(\mathbf p, \mathbf q)\Big)\Big( H_{\mathbf p}(\mathbf p, \mathbf q)+G(\mathbf p)\lambda\Big)= 0.\label{lamba}
\end{align}
Equation \eqref{lamba} enables the expression of $\lambda$ as a function of $(\mathbf p,\mathbf q)$.

Substituting  $\lambda(\mathbf p,\mathbf q)$  into \eqref{Hsys} yields a differential equation for coordinate $(\mathbf p,\mathbf q)$ on the following manifold:
\begin{align}
	\mathcal M=\{(\mathbf p,\mathbf q)|\phi(\mathbf p) = P, G(\mathbf p) H_{\mathbf q}(\mathbf p, \mathbf q)= 0\}
\end{align}

Denote $T^*\mathcal N=\{(\mathbf p, \mathbf q)|\mathbf p\in \mathcal N, \mathbf q \in T_{\mathbf p}^*\mathcal N\}$ as the cotangent bundle of $\mathcal N$, where $T_{\mathbf p}^*\mathcal N$ is the cotangent space for a fixed $\mathbf p\in \mathcal N$. The manifold $\mathcal M$ can be interpreted geometrically as the cotangent bundle of the configuration manifold $\mathcal N$, i.e., $\mathcal M=T^*\mathcal N$.
The constrained Hamiltonian system \eqref{Hsys}, with the Hamiltonian defined in \eqref{Hamilton}, can be interpreted as a differential equation on the cotangent bundle $\mathcal M$ of 
$\mathcal N$.

The preservation of the Hamiltonian for  \eqref{Hsys} can be easily proved by differentiate $H(\mathbf p,\mathbf q)$ with respect to
$t$ as
\begin{align}
	-H_{\mathbf q}(\mathbf p, \mathbf q)^TH_{\mathbf p}(\mathbf p, \mathbf q)-H_{\mathbf q}(\mathbf p, \mathbf q)^TG(\mathbf p)^T\lambda\notag\\
	+
	H_{\mathbf p}(\mathbf p, \mathbf q)^TH_{\mathbf q}(\mathbf p, \mathbf q)
\end{align}

The first and last terms cancel, while the central term disappears since 
$G(\mathbf p) H_{{\mathbf q}}(\mathbf p, \mathbf q)=0$. As a result, the Hamiltonian 
$H(\mathbf p, \mathbf q)$ remains constant along the trajectories defined by equation \eqref{Hsys}.
Moreover, the flow of \eqref{Hsys} represents a symplectic transformation on $\mathcal M$. The concise proof is  provided in \cite{symplectic3}.

As we have mentioned in the previous subsection, the augmented Hamlitonian is time independent, which means the system in \eqref{Hsys} is conservative. The evolution of the system takes place on the level sets of 
$H$ without reaching a state where the pontional energy is minimized \cite{jordandan}. To reach a point where the pontional energy is minimized, the dissipation is introduced, guiding the evolution towards a minimizer 
$\mathbf p^*, \mathbf q^*$ of the augmented Hamiltonian with dissipation.

Define  the augmented Hamiltonian with dissipation as \cite{jordandan}
\begin{align}
	{\tilde{H}}(\tilde{\mathbf p}, \tilde{\mathbf q})+\lambda\phi(\tilde{\mathbf p})=e^{\gamma t}\Big({{H}}(\tilde{\mathbf p}, e^{-\gamma t}\tilde{\mathbf q})+\lambda\phi(\tilde{\mathbf p})\Big)
\end{align}
where $(\tilde{\mathbf p}, \tilde{\mathbf q})=(\mathbf p,e^{\gamma t}\mathbf q)$.
The Hamiltonian equations for the augmented Hamiltonian with dissipation is given as
\begin{align}
	&\dot{\tilde{\mathbf p}}={\tilde{H}}_{\tilde{\mathbf q}}(\tilde{\mathbf p}, \tilde{\mathbf q})= H_{\mathbf q}(\mathbf p, \mathbf q)=\dot{\mathbf p}\notag\\ 
	&\dot{\tilde{\mathbf q}}=- {\tilde{H}}_{\tilde{\mathbf p}}(\tilde{\mathbf p}, \tilde{\mathbf q})-G(\tilde{\mathbf p})^T \lambda \nonumber \\
	& ~~=-e^{\gamma t}(H_{{\mathbf p}}({\mathbf p}, {\mathbf q})+G({\mathbf p})^T \lambda).
\end{align}
Since $\tilde{\mathbf q}=e^{\gamma t}\mathbf q$, we have $\dot{\tilde{\mathbf q}}=e^{\gamma t}(\dot{{\mathbf q}}+\gamma \mathbf q)$.
Then,  a dissipative version  of \eqref{Hsys} is given as
\begin{align}
	&\dot{\mathbf p}= H_{\mathbf q}(\mathbf p, \mathbf q)\notag\\ 
	&\dot{\mathbf q}=- H_{\mathbf p}(\mathbf p, \mathbf q)-G(\mathbf p)^T \lambda-\gamma \mathbf q\notag\\
	&\phi(\mathbf p)= P \label{dissa}
\end{align}
where $\gamma \textgreater 0$ denotes the coefficient of the dissipation.

\subsection{Numerical Method with RATTLE}

To solve the continuous dissipative version of Hamiltonian system \eqref{dissa},  a numerical scheme named RATTLE  integrator \cite{symplectic3} is used to discretize the system. 
First, we provide the gradient of $g(\mathbf p)$ which is needed in the numerical scheme.
Let $a_{k,c}$, $b_{k,c}$ and $c_{k,c}$ be defined as
\begin{align}
	a_{k,c}&=\Gamma_{k,c}^{-1}\mathbf h_{k,c}\mathbf p_k^c\\ b_{k,c}&=a_{k,c}c_{k,c}a_{k,c}^H\\
	c_{k,c}&=(1+(\mathbf p_k^c)^H\mathbf h_{k,c}^Ha_{k,c})^{-1}
\end{align}
Let the gradient of $f(\mathbf p)$ be defined as
\begin{align}
	{\rm {grad}}\, f(\mathbf p)=({\rm {grad}}\, f(\mathbf p_1^1),\cdots,{\rm {grad}}\, f(\mathbf p_1^{N_v}),\notag\\
	\cdots,{\rm {grad}}\, f(\mathbf p_K^1),\cdots,{\rm {grad}}\, f(\mathbf p_K^{N_v})).
\end{align}
where ${\rm {grad}}\, f(\mathbf p_k^c)=\sum_{l\neq k}^{K}w_lb_{l,c}\mathbf h_{l,c}^H\mathbf h_{l,c}\mathbf p_k^c-w_ka_{k,c}c_{k,c}\mathbf h_{k,c}^H$  is the $kc$-th component of ${\rm {grad}}\, f(\mathbf p)$.
For brevity, we define $m_{l,k,c}=w_lb_{l,c}\mathbf h_{l,c}\mathbf p_k^c$, $n_{k,c}=w_ka_{k,c}c_{k,c}$, and two diagonal matrices as $\mathbf A^l={\rm {Bdiag}}\{m_{l,1,1}\mathbf I,\cdots,m_{l,K,N_v}\mathbf I\}$ and  $\mathbf B={\rm {Bdiag}}\{n_{1,1}\mathbf I,\cdots,n_{K,N_v}\mathbf I\}$. Let $\mathbf h_l$ be defined as $\mathbf h_l=[\mathbf h_{l,1},\cdots,\mathbf h_{l,N_v}]^H$. Define vectors $\hat{\mathbf h}_l$ and $\mathbf h$ as $\hat{\mathbf h}_l=[\mathbf h_{l},\cdots,\mathbf h_{l}]^H$ and $\mathbf h=[\mathbf h_{1},\cdots,\mathbf h_{K}]^H$. Based on the above definitions, we can obtain Theorem 1.

{\itshape Theorem 1:} 
	The gradient of $g(\mathbf p)$  is obtained as
	\begin{align}
		{\rm {grad}}\, g(\mathbf p)=\sum_{l\neq k}^{K} \mathbf A^l \hat{\mathbf h}_l-\mathbf B \mathbf h+\alpha\mathbf H^H\mathbf M^H \mathbf M \mathbf H \mathbf p \label{gragg}
	\end{align}
The detailed proof is in Appendix A.

Since the  Hamiltonian is separable, $H_{\mathbf q}$ and $H_{\mathbf p}$ can be calculated as
$H_{\mathbf q}=\frac{1}{2}\mathbf M \mathbf q$ and
$H_{\mathbf p}={\rm {grad}}\, g(\mathbf p)$, respectively.
By using the RATTLE  integrator introduced in \cite{symplectic3} and assuming that $\mathbf M=\mathbf I$, the update of the position coordinate $\mathbf p$ and the momentum coordinate $\mathbf q$ can be obtained, and we can obtain Theorem 2.

{\itshape Theorem 2:} 
	The sequences $\mathbf p_n$ and $\mathbf q_n$ of dissipative system \eqref{dissa}  can be recursively updated as 
	\begin{align}
		\mathbf q_{n+1/2}&=e^{-\gamma h/2}\mathbf q_{n}-\frac{h}{2}(\sum_{l\neq k}^{K} \mathbf A^l_n \hat{\mathbf h}_l-\mathbf B_n \mathbf h\notag\\
		&+\alpha\mathbf H^H \mathbf H \mathbf p_n+\lambda_n \mathbf p_{n})\label{q12} \\
		\mathbf p_{n+1}&=\mathbf p_{n}+\frac{h}{2} \mathbf q_{n+1/2}\label{pp}\\
		\mathbf q_{n+1}&=e^{-\gamma h/2}(\mathbf q_{n+1/2}-\frac{h}{2}(\sum_{l\neq k}^{K} \mathbf A^l_{n+1} \hat{\mathbf h}_l-\mathbf B_{n+1} \mathbf h\notag\\
		&\,\,\,\,\,\,+\alpha\mathbf H^H \mathbf H \mathbf p_{n+1}+\mu_n \mathbf p_{n+1}))\label{qq}
	\end{align}
	where $h$ is the step length and the parameter $\mu_n$ is introduced to ensure that
	$\mathbf q_{n+1}$ resides in $T_{\mathbf p_{n+1}}^*\mathcal N$. The calculation of $\lambda_n$ and $\mu_n$ can be given as 
	\begin{align}
		\lambda_n&=\frac{\mathbf q_n^H\mathbf q_n-2\mathbf p_n^H ({\rm {grad}}\, g(\mathbf p_{n}))}{2\mathbf p^H\mathbf p}\label{lamb}\\
		\mu_n&=\frac{2\mathbf p_{n+1}^H\mathbf q_{n+1/2}-h\mathbf p_{n+1}^H({\rm {grad}}\, g(\mathbf p_{n+1}))}{h\mathbf p_{n+1}^H\mathbf p_{n+1} }.\label{mun}
	\end{align}
	The sequences can converge to an local optimum of the problem \eqref{problem_qos}.
	The detailed proof is in Appendix B.

In the following, we  develop a  scheme that adaptively adjusts the step length 
$h$ to accelerate the optimization process. Rather than employing a traditional line search or trust region approach, we adopt the method proposed in \cite{step}, where the optimization algorithm is modeled as a continuous dynamical system. The step size is dynamically adjusted to ensure efficient time discretization of this system, thereby improving the optimization efficiency.

Specifically, we employ a proportional controller (P control) to adjust $h$ based on a parameter $\delta_n$, which can be calculated by referring to \cite{step}. P control is a basic form of the broader PID (Proportional-Integral-Derivative) control family, which is widely used for automated regulation of user-defined quantities in dynamic systems. In this case, the target quantity for control is $\delta_n$. While the use of PID control for step length adjustment in discretization schemes for differential equations has been explored previously, the novelty of following approach lies in applying this concept to optimization.

The P control method introduces two key parameters: 
$r$, representing the target error, and 
$\theta$, which  can be interpreted as a gain factor. This control method provides a systematic approach for adjusting the step length, based on these parameters, to modulate the update process efficiently.
\begin{align}
	h_{n+1}=(\frac{r}{\delta_n})^{\theta/2}h_{n}.\label{stepl}
\end{align}

After taking logarithms, the equation \eqref{stepl} can be written as
\begin{align}
	\log h_{n+1}=\log h_{n}+\frac{\theta}{2}(\log r-\log \delta_n).\label{step}
\end{align}
According to the derivatives in \cite{step}, the equation \eqref{step} can be further written as
\begin{align}
	\log h_{n+1}=(1-\theta)\log h_{n}+\frac{\theta}{2}(\log r-\log C)
\end{align}
where $C$ is  a constant when  $h_{n}$ is small.

The logarithm of the step length, $\log h_{n}$, approximately follows a linear system and is anticipated to converge to its steady-state, where 
$\delta_n=r$, at a convergence rate of $1-\theta$ \cite{step}.
Regarding parameter selection, 
$\theta$ must lie within the range 
$[0,2]$\cite{steprange}, with 
$\theta=0$ corresponding to a constant step length. As 
$\theta$ increases, greater variations in 
$h$ are permitted,  reducing the number of steps required for convergence. The controller's performance is notably sensitive to changes in 
$\theta$, while it shows less sensitivity to variations in 
$r$. More specific analysis is presented in \cite{step}.
Based on the adaptive step length, algorithm 1 summarizes
the proposed CSPD.

\begin{algorithm}[htbp]
	\caption{ CSPD  with symplectic optimization}
	\begin{enumerate}[\IEEElabelindent=3em]
		\item[ 1:]
		Set iteration $n=0$. Initialize $h_n$,  $\mathbf p_n$ and $\mathbf q_n$
		\item[ 2:]
		Calculate $\mathbf q_{n+1/2}$  as
		\begin{align}
			\mathbf q_{n+1/2}&=e^{-\gamma h\/2}\mathbf q_{n}-\frac{h_n}{2}(\sum_{l\neq k}^{K} \mathbf A^l_n \hat{\mathbf h}_l-\mathbf B_n \mathbf h\notag\\
			&+\alpha\mathbf H^H \mathbf H \mathbf p_n+\lambda_n \!\mathbf p_{n})\notag
		\end{align} where $\lambda_n$ can be calculated according to \eqref{lamba}
		\item[3:]
		Update the position coordinate $\mathbf p_{n+1}$ as	
		\begin{align}
			\mathbf p_{n+1}=\mathbf p_{n}+\frac{h_n}{2} \mathbf q_{n+1/2}\notag
		\end{align}
		\item[ 4:]
		Update ${\rm {grad}}\, g(\mathbf p_{n+1})$ according to Theorem 1.
		\item[ 5:]
		Update the momenta coordinate $\mathbf q_{n+1}$  as
		\begin{align}
			\mathbf q_{n+1}&=e^{-\gamma h/2}(\mathbf q_{n+1/2}-\frac{h}{2}(\sum_{l\neq k}^{K} \mathbf A^l_{n+1} \hat{\mathbf h}_l\notag\\
			&-\mathbf B_{n+1} \mathbf h+\alpha\mathbf H^H \mathbf H \mathbf p_{n+1}+\mu_n \mathbf p_{n+1}))\notag
		\end{align}	where $\mu_n$ is given in \eqref{mun}.
		\item[6:]	
		Update the step length as
		\begin{align}
			h_{n+1}=(\frac{r}{\delta_n})^{\theta/2}h_{n}\notag
		\end{align}	
		Update the iteration as $n=n+1$.
			
		Repeat Steps 2 to 6 until the	preset goal is reached or until convergence.
	\end{enumerate}
\end{algorithm}

The computational complexity of  CSPD  with symplectic optimization is primarily determined by calculating ${\rm {grad}}\, g(\mathbf p)$, which can be given as $O(N_S(K^2N_vM_t+KN_vM_t))$  where $N_S$ is the total iteration number of Algorithm 1. We choose the WMMSE precoder design  in \cite{christensen_weighted_2008} for comparison. The complexity of WMMSE precoder design is $O(N_W(K^2N_vM_t+KN_vM_t))$, where $N_W$ is the total iteration number of WMMSE.

\section{SIMULATION RESULTS}

\subsection{Simulation Setups}

Simulation results are presented to evaluate the performance of the proposed  CSPD. 
The QuaDriGa \cite{jaeckel_quadriga_2014}, which employs a geometry-based stochastic channel model (GSCM), is utilized to evaluate the posteriori beam-based statistical channel model.
The channels generated by QuaDriGa effectively capture the statistical parameters of the beam-based model and approximate the physical channel.  
The simulation scenario follows "3GPP\_3D\_UMa\_NLOS",  with the antenna array set as "3gpp-3d" and the UPA antenna configuration specified as  $M_t = 128$, where $M_x = 16$ and $M_z = 8$. Users are randomly distributed around the BS, and    $N_v=256$ subcarriers are used for transmission.

\subsection{Comparisons with per-subcarrier precoder design}

The  WSR performance of the proposed CSPD is evaluated  in comparison with the per-subcarrier precoder design. The per-subcarrier precoder utilizes the conventional WMMSE precoder, which requires computationally expensive matrix inversion operations in massive MIMO-OFDM systems. To mitigate this computational burden, the conjugate gradient (CG) method can be employed as an alternative to avoid matrix inversion. We combine the WMMSE precoder with the CG method as a baseline for comparison. Simulation results demonstrate that, with 40 iterations, the proposed CSPD without channel smoothing achieves slightly better performance than that with channel smoothing, while both outperform the WMMSE precoder design. Furthermore, when the number of WMMSE iterations is increased to 150, its performance matches that of the CSPD without channel smoothing. These results highlight the efficiency of the symplectic optimization algorithm, which significantly reduces the required number of iterations and accelerates convergence.

\begin{figure}[h]
	
	\centering
	
	\includegraphics[width=0.5\textwidth]{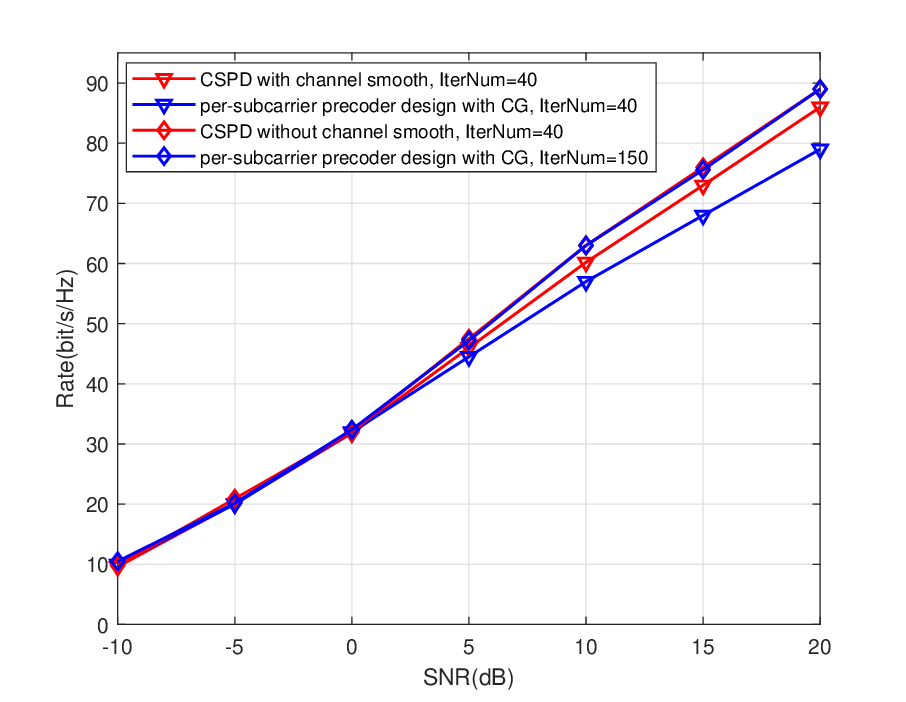}
	
	\caption{The comparisons for CSPD and per-subcarrier precoder design on the WSR performance metric  with $M_t = 128$ and $K=10$.}
	
	\label{fig:comp_approx1}
\end{figure}

\begin{figure}[h]
	
	\centering
	
	\includegraphics[width=0.5\textwidth]{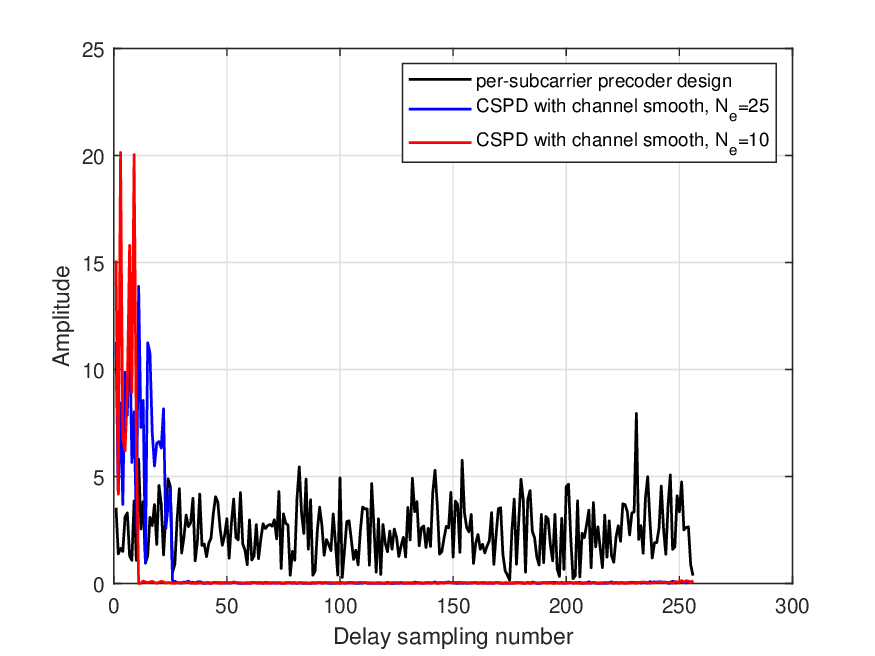}
	
	\caption{The delay domain effective channel of CSPD and per-subcarrier precoder design.}
	
	\label{fig:Delay}
\end{figure}

Then, we transform
the frequency domain effective channel after precoding  into the delay domain. From  Fig.~\ref{fig:Delay}, it is observed that the proposed CSPD, which considers effective channel smoothing, can effectively reduce the delay spread of the effective channel  compared to the conventional per-subcarrier precoder design. Moreover, the proposed CSPD with effective channel smoothing allows for the adjustment of the delay spread through appropriate adjustment of the parameter $N_e$. The delay spread of the delay domain effective channels becomes smaller, i.e., the correlation between frequency domain effective channels  becomes larger, which facilitates effective channel estimation and signal detection.

%
%
%
%
%
%
%
%

\subsection{Effective Channel Estimation Performance}
We discuss the
effective channel estimation and signal detection when the proposed transform-based CSPD is used.
Consider the MMSE channel estimation at the user ends.
For simplicity, the subscript $k$ is dropped in the following derivation.
Let $y_{c}$ and $\bar{h}_c=\mathbf h_{c} \mathbf p_c$ denote the received signal and downlink effective channel at the user side for $c$-th subcarrier.
Define $\mathbf y=[y_{1},y_{2},\cdots,y_{N_v}]^T$, $\mathbf h_f=[\bar{h}_1,\bar{h}_2,\cdots,\bar{h}_{N_v}]^T$, and $\mathbf z=[z_{1},z_{2},\cdots,z_{N_v}]^T$,
the transmission model for downlink effective channel estimation is given as
\begin{align}
	\mathbf y
	= \mathbf X \mathbf h_f+ \mathbf z
\end{align}
where $\mathbf X={\rm Diag}\{x_1,x_2,\cdots,x_{N_v}\}$ is the  pilot signal whose elements can be provided based on the Zadoff-Chu (ZC) sequence. Let $N_d$ denote the delay sampling number.
The  delay domain channel can be transformed to frequency domain channel by
\begin{align}
	\mathbf h_f=\mathbf U_f \mathbf h_d
\end{align}
where $\mathbf U_f \in \mathbb{C}^{N_v \times N_d}$ is a partial DFT and $\mathbf h_d \in \mathbb{C}^{N_d \times 1}$ is the delay domain channel.
Further, the delay domain channel can be decomposed as
\begin{align}
	\mathbf h_d=\tilde{\mathbf m}\odot \tilde{\mathbf w},
\end{align}
where  $\tilde{\mathbf m}$ is a 
deterministic vector and  $\tilde{\mathbf w}$ consists of i.i.d. entries following $\mathcal{CN}(0, 1 )$ distribution.
Let $\tilde{\boldsymbol{\omega}}=\tilde{\mathbf m}\odot \tilde{\mathbf m}$, we have $\mathbb{E}\{\mathbf h_d\mathbf h_d^H\}=\boldsymbol{\Lambda}$ where the diagonal matrix $\boldsymbol{\Lambda}_{ii}=\tilde{\boldsymbol{\omega}}_{i}$.

\begin{figure}[h]
	
	\centering
	
	\includegraphics[width=0.5\textwidth]{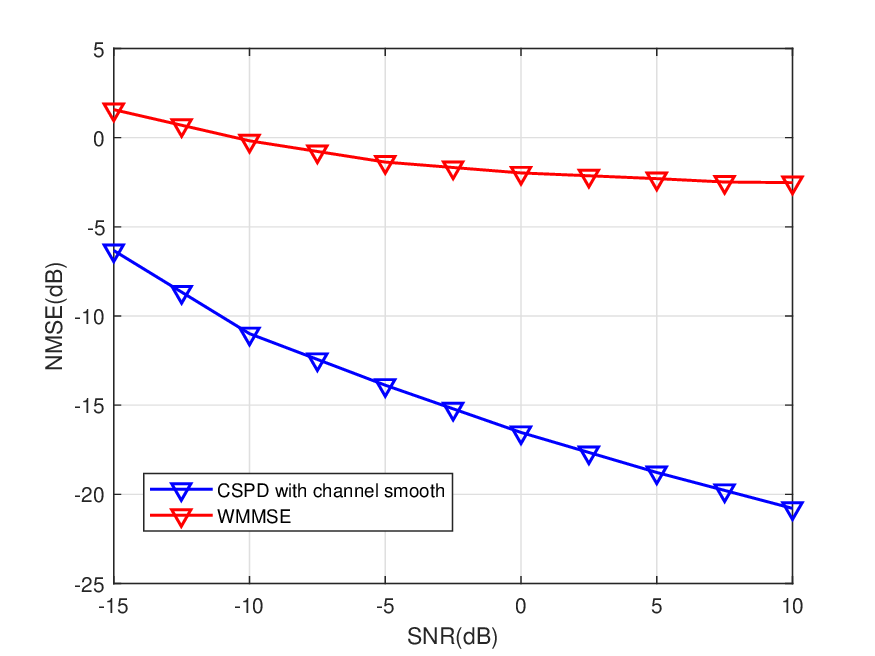}
	
	\caption{The NMSE performance of the CSPD
		compared with the WMMSE precoder design.}
	
	\label{fig:ce}
	
\end{figure}

Let $\mathbf R_{\mathbf h_f}=\mathbb{E}\{\mathbf h_f\mathbf h_f^H\}$, we have $\mathbf R_{\mathbf h_f}=\mathbf U_f\boldsymbol{\Lambda} \mathbf U_f^H$. By using the MMSE channel estimation,  the estimated frequency domain channel is given as
\begin{align}
	\hat{\mathbf h}_{f,mmse}&=\mathbf R_{\mathbf h_f}(\mathbf X^H \mathbf X\mathbf R_{\mathbf h_f}+\delta_z^2 \mathbf I)^{-1}\mathbf X^H\mathbf y
	\notag\\
	&=\mathbf U_f\boldsymbol{\Lambda} \mathbf U_f^H(\mathbf X^H\mathbf X\mathbf U_f\boldsymbol{\Lambda} \mathbf U_f^H+\delta_z^2 \mathbf I)^{-1}\mathbf X^H\mathbf y.
\end{align}
The normalized mean squared error (NMSE) performance of the channel estimation in the  proposed CSPD case and the  per-subcarrier precoder design case is given in Fig.~\ref{fig:ce}.   We utilize comb-type pilots for channel estimation with a pilot interval set to $2$. The result shows that the effective channels can be estimated more precisely when the proposed CSPD is used. The significant channel estimation performance improvement comes from the enhanced  smoothness of the frequency domain effective channel by using the proposed precoders.

Then, we perform the signal detection. 
With the definition of ${{ z}}_{k,c}^{'}$  and  the covariance  $\Gamma_{k,c}$ provided in Section II, 
the received signal is rewritten as
\begin{align}
	y_{k,c}={\hat h}_{k,c}  x_{k,c} +  {{ z}}_{k,c}^{'} \label{dec_receive}
\end{align}
where ${\hat h}_{k,c}$ represents the estimated effective channel.

Consider the MMSE detection, the estimated signal is given as
\begin{align}
	\hat{x}_{k,c}=({\hat h}_{k,c}^*{\hat h}_{k,c}+\Gamma_{k,c})^{-1}{\hat h}_{k,c}^* y_{k,c}. \label{dec_MMSE}
\end{align}

\begin{figure}[h]
	
	\centering
	
	\includegraphics[width=0.5\textwidth]{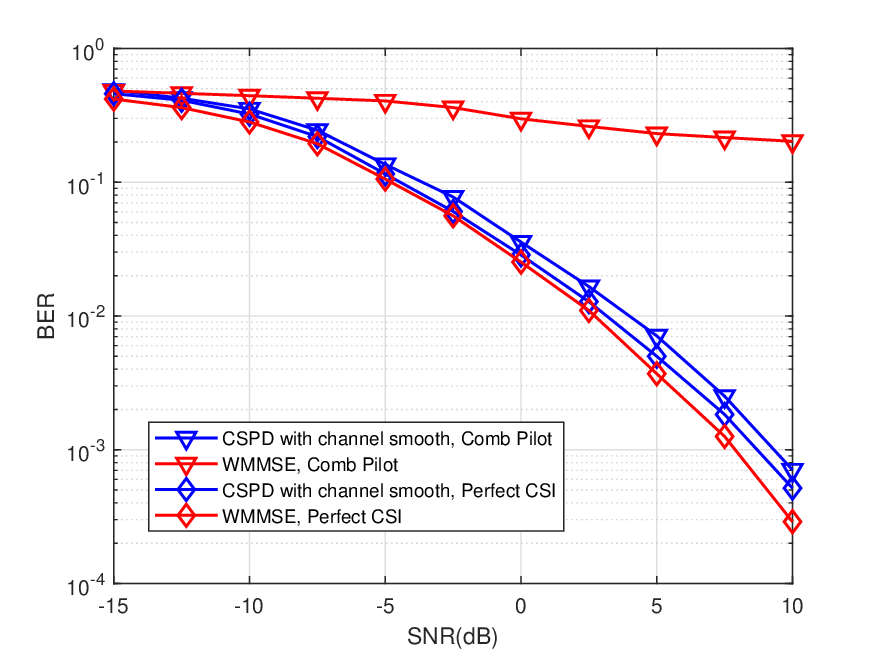}
	
	\caption{The BER performance of the CSPD
		compared with the the WMMSE precoder design.}
	
	\label{fig:detect}
	
\end{figure}

The modulation scheme is the quadrature phase shift keying. The  bit error rate (BER) performance of the proposed CSPD and the  per-subcarrier precoder design is given in Fig.~\ref{fig:detect}. 	The
BER performance of the proposed CSPD outperforms that of the  per-subcarrier precoder design, primarily due to the improved effective channel estimation performance.

%
%
%
%
%
%

\subsection{Convergence Analysis}

\begin{figure}[h]
	
	\centering
	
	\includegraphics[width=0.5\textwidth]{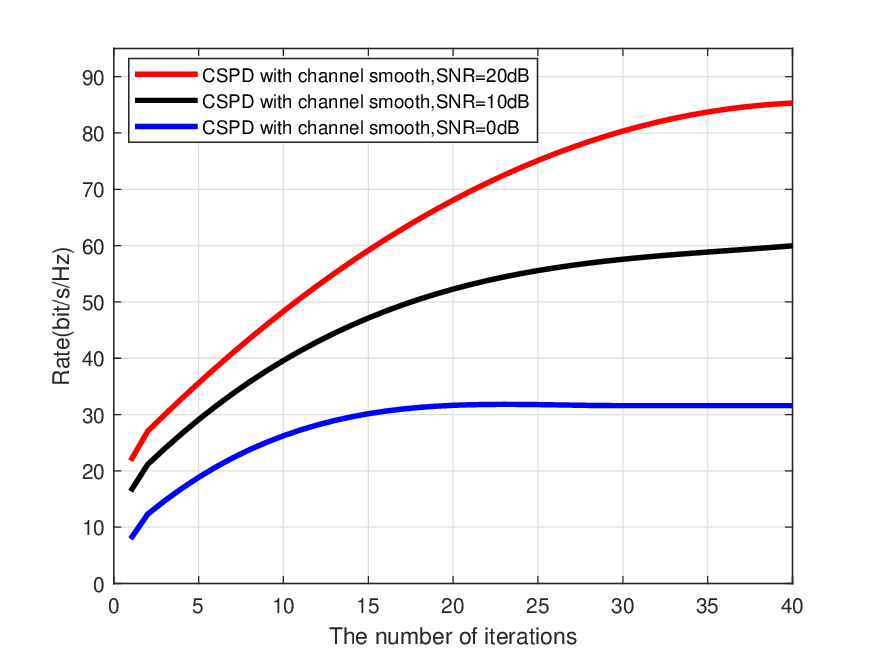}
	
	\caption{The convergency of the  CSPD.}
	\label{fig:convergen}
	
\end{figure}

In Fig.~\ref{fig:convergen}, the WSR performance of the  CSPD is depicted for different SNRs against the iteration count. We conclude from Fig.~\ref{fig:convergen} that the objective function values of the proposed algorithm quickly converges at low SNRs.
More iterations are required for convergence as SNR increases.  Specifically,  $20$ iterations are sufficient for the convergence of the proposed  CSPD   at  SNR$=0$dB, whereas $40$ iterations are needed as the SNR increases to $20$dB.

\begin{figure}[h]
	
	\centering
	
	\includegraphics[width=0.5\textwidth]{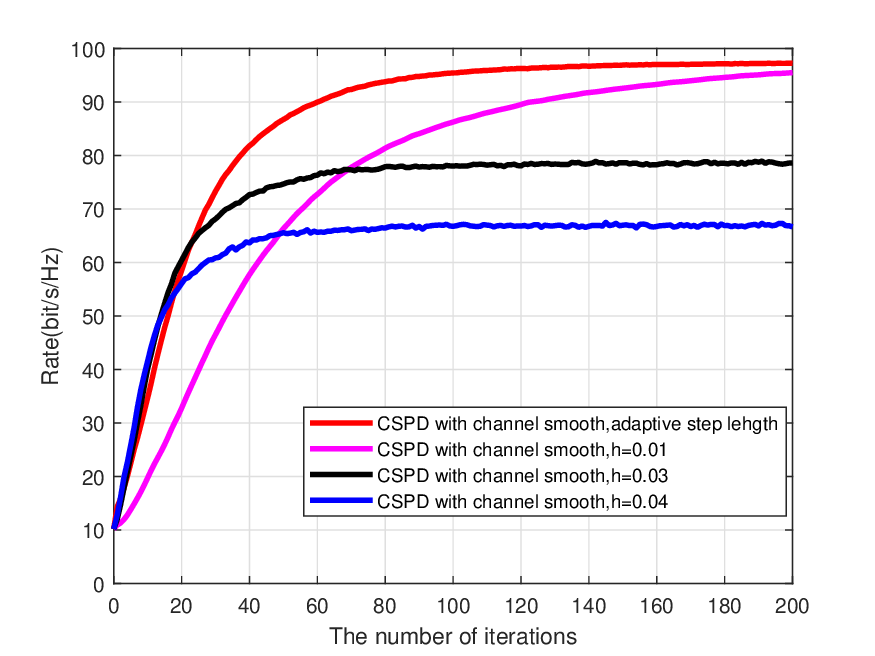}
	
	\caption{The effect of step length on convergence performance.}
	\label{fig:step}
	
\end{figure}

Next, we compare the convergence of the proposed algorithm with adaptive step length and fixed step length. In figure.~\ref{fig:step}, the WSR performance of the  CSPD is depicted for different step length schemes against the iteration count. For the CSPD with fixed step length, if the step length is set too small, although it can converge to a point with better performance, the number of convergences required is very large. If the step length is set too large, it can converge faster, but the performance of the converged point will be relatively poor. Compared to the CSPD with fixed step length, the CSPD with adaptive step length can converge to a point with better performance faster.

\section{Conclusion}

In this paper, we propose a CSPD for massive MIMO-OFDM systems that jointly optimizes the WSR performance and the smoothness of the effective channel. By introducing a delay indicator function to characterize the impact of the large delay components in the delay domain effective channel, we formulate an optimization problem with weighted sum of two objectives to balance the WSR and the  delay indicator function.  The delay spread of effective channel can be reduced by appropriately adjusting the parameters in the objective function. Then,
we  introduce a dissipative augmented Hamiltonian system related to the optimization problem and  use the symplectic integration to obtain a stable algorithm by discretizing the Hamiltonian system. The simulation results shows that the proposed CSPD provides satisfying WSR performance while maintaining channel smoothness.

\appendices
\section{Proof of Theorem 1}

The  gradient of $ g(\mathbf p)$ can be written as
\begin{align}
	{\rm {grad}}\, g(\mathbf p)={\rm {grad}}\, f(\mathbf p)+\alpha({\rm {grad}}\, d(\mathbf p))\label{a1}
\end{align}
where ${\rm {grad}}\, f(\mathbf p)$ is constructed by stacking the gradient ${\rm {grad}}\, f(\mathbf p_k^c)$ for each user and subcarrier as 
\begin{align}
	{\rm {grad}}\, f(\mathbf p)&=({\rm {grad}}\, f(\mathbf p_1^1),\cdots,{\rm {grad}}\, f(\mathbf p_1^{N_v})\notag\\
	&,\cdots,{\rm {grad}}\, f(\mathbf p_K^1),\cdots,{\rm {grad}}\, f(\mathbf p_K^{N_v})).
\end{align}

We calculate the derivation of ${\mathcal{R}_{k,c}}$ and ${\mathcal{R}_{l,c}}$ with respect to $(\mathbf p_k^c)^*$ as 
\begin{align}
	\frac{\partial {\mathcal{R}_{k,c}}}{\partial (\mathbf p_k^c)^*}&=\frac{1}{1+\Gamma_{k,c}^{-1}\mathbf h_{k,c}\mathbf p_k^c(\mathbf p_k^c)^H\mathbf h_{k,c}^H} \mathbf h_{k,c}^H\Gamma_{k,c}^{-1}\mathbf h_{k,c}\mathbf p_k^c\notag\\
	&=a_{k,c}c_{k,c}\mathbf h_{k,c}^H
\end{align}
\begin{align}
	\frac{\partial {\mathcal{R}_{l,c}}}{\partial (\mathbf p_k^c)^*}&=[(\Gamma_{l,c}+\mathbf h_{l,c}\mathbf p_l^c(\mathbf p_l^c)^H\mathbf h_{l,c}^H)^{-1}-\Gamma_{l,c}^{-1}]\mathbf h_{l,c}^H\mathbf h_{l,c}\mathbf p_k^c
	\notag\\
	&=-b_{l,c}\mathbf h_{l,c}^H\mathbf h_{l,c}\mathbf p_k^c
\end{align}
Since $f(\mathbf p_1,\cdots,\mathbf p_K)=-\sum\limits_{k = 1}^K\sum\limits_{c = 1}^{N_v}w_k {{\mathcal{R}_{k,c}}}$, we have
\begin{align}
	{\rm {grad}}\, f(\mathbf p_k^c)=-w_ka_{k,c}c_{k,c}\mathbf h_{k,c}^H+\sum\limits_{l\neq k}^{K}w_lb_{l,c}\mathbf h_{l,c}^H\mathbf h_{l,c}\mathbf p_k^c.\label{a2}
\end{align}
Further, ${\rm {grad}}\, \phi_2(\mathbf p)$ can be easily calculated as
\begin{align}
	{\rm {grad}}\, \phi_2(\mathbf p)=\mathbf H^H\mathbf M^H \mathbf M \mathbf H \mathbf p.\label{a3}
\end{align}
Based on the definitions of $\mathbf A_l$, $\hat{\mathbf h}_l$ and $\mathbf h$, we combine \eqref{a1}, \eqref{a2} and \eqref{a3} to obtain the Theorem 1.

\section{Proof of Theorem 2}
Conservative system: The conservative system corresponds to equation \eqref{Hsys}. We denote its flow by 
$\Phi_t^C$, and to approximate this flow by $\Phi_h^C$, we apply the RATTLE numerical method. Let 
$h$ represent the  step length, and 
$(\mathbf p_n,\mathbf q_n)$ be the approximation of the flow at time $t_n$. To compute the approximation at 
$t_{n+1}$, the RATTLE algorithm can be extended to general Hamiltonians
as follows according to \cite{symplectic3,ANDERSEN198324}

\begin{align}
	\mathbf q_{n+1/2}=\mathbf q_{n}-\frac{h}{2}(H_{\mathbf p}(\mathbf p_n)+G(\mathbf p_n)\lambda_n),\notag\\
	\mathbf p_{n+1}=\mathbf p_{n}+hH_{\mathbf q}(\mathbf q_{n+1/2})\notag\\
	0=\phi_1(\mathbf p_{n+1})\notag\\
	\mathbf q_{n+1}=\mathbf q_{n+1/2}-\frac{h}{2}(H_{\mathbf p}(\mathbf p_{n+1})+G(\mathbf p_{n+1})\mu_n)\notag\\
	0=G(\mathbf p_{n+1})^HH_{\mathbf q}(\mathbf q_{n+1})
\end{align}

This method guarantees that 
$\Phi_h^C(\mathbf p_n,\mathbf q_n)=(\mathbf p_{n+1},\mathbf q_{n+1})$ remains constrained on $\mathcal{M}$. 
Specifically, the parameter $\lambda_n$ and the equation $0=\phi_1(\mathbf p_{n+1})$, ensures that the constraint is
satisfied, thereby placing $\mathbf p_{n+1}$ on
$\mathcal N$. Similarly, the parameter $\mu_n$ and the  equation $0=G(\mathbf p_{n+1})^HH_{\mathbf q}(\mathbf q_{n+1})$ guarantee  that
$\mathbf q_{n+1}$ resides in $T_{\mathbf p_{n+1}}^*\mathcal N$, as $\mathbf q_{n+1}$
is effectively the projection of 
$\mathbf q_{n+1/2}$ onto $T_{\mathbf p_{n+1}}^*\mathcal N$. The existence and local uniqueness of the solution are discussed in \cite{jordandan}.
However, the system remains nonlinear and implicit due to the nonlinearity of $g$ and the unknown nature of $\lambda_n$, as its computation requires $(\mathbf p_{n}, \mathbf q_{n+1/2})$.

Dissipative system: The dissipative component of the system \eqref{Hsys} is
\begin{align}
	\dot{\mathbf p}=\mathbf 0\\
	\dot{\mathbf q}=-\gamma \mathbf q,
\end{align}
which is straightforward to solve exactly. Given the initial point $(\mathbf p_{0}, \mathbf q_{0}) \in \mathcal M$, the flow 
$\Phi_t^D$ at time $t$ is presented as
\begin{align}
	\Phi_t^D(\mathbf p_{0}, \mathbf q_{0})=(\mathbf p_{0}, e^{-\gamma t}\mathbf q_{0}).
\end{align}

Splitting scheme: The overall flow of system \eqref{Hsys} is determined by composing the conservative flow $\Phi_t^C$ with the dissipative flow $\Phi_t^D$.
\begin{align}
	\Phi_h=\Phi_{h/2}^D\circ\Phi_h^C\circ\Phi_{h/2}^D.
\end{align}
Then, the leapfrog symmetric second order integration scheme is used  to obtain the final numerical method according to \cite{jordandan}
\begin{align}
	\mathbf q_{n+1/2}=e^{-\gamma h/2}\mathbf q_{n}-\frac{h}{2}(H_{\mathbf p}(\mathbf p_n)+G(\mathbf p_n)\lambda_n),\notag\\
	\mathbf p_{n+1}=\mathbf p_{n}+hH_{\mathbf q}(\mathbf q_{n+1/2})\notag\\
	0=\phi_1(\mathbf p_{n+1})\notag\\
	\mathbf q_{n+1}=e^{-\gamma h/2}(\mathbf q_{n+1/2}-\frac{h}{2}(H_{\mathbf p}(\mathbf p_{n+1})+G(\mathbf p_{n+1})\mu_n))\notag\\
	0=G(\mathbf p_{n+1})^HH_{\mathbf q}(\mathbf q_{n+1})\label{sequen}
\end{align}
By substituting $H_{\mathbf q}=\frac{1}{2}\mathbf M \mathbf q$ and
$H_{\mathbf p}={\rm {grad}}\, g(\mathbf p)$ into \eqref{sequen}, we obtain the updated sequences $\mathbf p_n$ and $\mathbf q_n$ in Theorem 2.
To satisfy the two constraints $\phi(\mathbf p_{n+1})=P$ and $G(\mathbf p_{n+1})^HH_{\mathbf q}(\mathbf q_{n+1})=0$,
The calculation of $\lambda_n$ can be obtained according to \eqref{lamba}. Let $\mathbf M=\mathbf I$. The equation \eqref{lamba} can be written as
\begin{align}
	\frac{1}{2}\mathbf q_n^H\mathbf q_n=\mathbf p_n^H ({\rm {grad}}\, g(\mathbf p_n)) +\lambda \mathbf p_n^H\mathbf p_n.
\end{align}
Then, we can obtain the calculation of $\lambda_n$ in Theorem 2.
The calculation of $\mu_n$ can be obtained by substituting $\mathbf q_{n+1}$ into $G(\mathbf p_{n+1})^HH_{\mathbf q}(\mathbf q_{n+1})=0$ as
\begin{align}
	\mathbf p_{n+1}^H(\mathbf q_{n+1/2}-\frac{h}{2}({\rm {grad}}\, g(\mathbf p_(n+1)) -\frac{h}{2} \mu_n \mathbf p_{n+1})=0.
\end{align}
Then, we can obtain the calculation of $\mu_n$ in Theorem 2.

\bibliographystyle{IEEEtran}
\bibliography{reference}

\begin{thebibliography}{10}
\providecommand{\url}[1]{#1}
\csname url@samestyle\endcsname
\providecommand{\newblock}{\relax}
\providecommand{\bibinfo}[2]{#2}
\providecommand{\BIBentrySTDinterwordspacing}{\spaceskip=0pt\relax}
\providecommand{\BIBentryALTinterwordstretchfactor}{4}
\providecommand{\BIBentryALTinterwordspacing}{\spaceskip=\fontdimen2\font plus
\BIBentryALTinterwordstretchfactor\fontdimen3\font minus
  \fontdimen4\font\relax}
\providecommand{\BIBforeignlanguage}[2]{{%
\expandafter\ifx\csname l@#1\endcsname\relax
\typeout{** WARNING: IEEEtran.bst: No hyphenation pattern has been}%
\typeout{** loaded for the language `#1'. Using the pattern for}%
\typeout{** the default language instead.}%
\else
\language=\csname l@#1\endcsname
\fi
#2}}
\providecommand{\BIBdecl}{\relax}
\BIBdecl

\bibitem{8804165}
M.~A. Albreem, M.~Juntti, and S.~Shahabuddin, ``Massive {MIMO} detection
  techniques: {A} survey,'' \emph{IEEE Commun. Surv. Tutor.}, vol.~21, no.~4,
  pp. 3109--3132, 2019.

\bibitem{8861014}
L.~Sanguinetti, E.~Björnson, and J.~Hoydis, ``Toward massive {MIMO} 2.0:
  Understanding spatial correlation, interference suppression, and pilot
  contamination,'' \emph{IEEE Trans. Commun.}, vol.~68, no.~1, pp. 232--257,
  2020.

\bibitem{9003618}
S.~Chen, Y.-C. Liang, S.~Sun, S.~Kang, W.~Cheng, and M.~Peng, ``Vision,
  requirements, and technology trend of {6G}: {How} to tackle the challenges of
  system coverage, capacity, user data-rate and movement speed,'' \emph{IEEE
  Wireless Commun.}, vol.~27, no.~2, pp. 218--228, 2020.

\bibitem{10026502}
X.~Liu, W.~Wang, X.~Gong, X.~Fu, {X. Q. Gao}, and X.-G. Xia, ``Structured
  hybrid message passing based channel estimation for massive {MIMO-OFDM}
  systems,'' \emph{IEEE Trans. Veh. Technol.}, vol.~72, no.~6, pp. 7491--7507,
  2023.

\bibitem{9364875}
C.~Wu, X.~Yi, W.~Wang, L.~You, Q.~Huang, {X. Q. Gao}, and Q.~Liu, ``Learning to
  localize: {A} {3D} {CNN} approach to user positioning in massive {MIMO-OFDM}
  systems,'' \emph{IEEE Trans. Wireless Commun.}, vol.~20, no.~7, pp.
  4556--4570, 2021.

\bibitem{8753608}
Y.~Ge, W.~Zhang, F.~Gao, S.~Zhang, and X.~Ma, ``Beamforming network
  optimization for reducing channel time variation in high-mobility massive
  {MIMO},'' \emph{IEEE Trans. Commun.}, vol.~67, no.~10, pp. 6781--6795, 2019.

\bibitem{luo}
X.~Zhao, S.~Lu, Q.~Shi, and Z.-Q. Luo, ``Rethinking {WMMSE}: {Can} its
  complexity scale linearly with the number of {BS} antennas?'' \emph{IEEE
  Trans. Signal Process.}, vol.~71, pp. 433--446, 2023.

\bibitem{ge}
K.~Chen, J.~Yang, Q.~Li, and X.~Ge, ``Sub-array hybrid precoding for massive
  {MIMO} systems: {A} {CNN}-based approach,'' \emph{IEEE Commun. Lett.},
  vol.~25, no.~1, pp. 191--195, 2021.

\bibitem{zhang}
Y.~Zhang, A.-A. Lu, B.~Liu, {X. Q. Gao}, and X.-G. Xia, ``Cross-subcarrier
  precoder design for massive {MIMO-OFDM} downlink,'' \emph{in Proc. IEEE 98th
  Veh. Technol. Conf.}, pp. 1--6, 2023.

\bibitem{zhang_sum-rate-optimal_2022}
Y.-X. Zhang, A.-A. Lu, and {X. Q. Gao},
  ``\BIBforeignlanguage{en}{{Sum}-rate-optimal statistical precoding for {FDD}
  massive {MIMO} downlink with deterministic equivalents},''
  \emph{\BIBforeignlanguage{en}{IEEE Trans. Veh. Technol.}}, vol.~71, no.~7,
  pp. 7359--7370, Jul. 2022.

\bibitem{1683918}
H.~Weingarten, Y.~Steinberg, and S.~Shamai, ``The capacity region of the
  gaussian multiple-input multiple-output broadcast channel,'' \emph{IEEE
  Trans. Inform. Theory.}, vol.~52, no.~9, pp. 3936--3964, 2006.

\bibitem{6812124}
A.~Kammoun, A.~Müller, E.~Björnson, and M.~Debbah, ``Linear precoding based
  on polynomial expansion: Large-scale multi-cell {MIMO} systems,'' \emph{IEEE
  J. Sel. Topics Signal Process.}, vol.~8, no.~5, pp. 861--875, 2014.

\bibitem{8668481}
L.~D. Nguyen, H.~D. Tuan, T.~Q. Duong, and H.~V. Poor, ``Multi-user regularized
  zero-forcing beamforming,'' \emph{IEEE Trans. Signal Process.}, vol.~67,
  no.~11, pp. 2839--2853, 2019.

\bibitem{1046557}
M.~Joham, K.~Kusume, W.~Utschick, and J.~Nossek, ``Transmit matched filter and
  transmit wiener filter for the downlink of {FDD DS-CDMA} systems,'' \emph{in
  Proc. IEEE 7th Symp. Spread-Spectrum Technol.}, vol.~5, pp. 2312--2316 vol.5,
  2002.

\bibitem{zhang_sum-rate_2017}
C.~Zhang, Y.~Huang, Y.~Jing, S.~Jin, and L.~Yang,
  ``\BIBforeignlanguage{en}{{Sum}-rate analysis for massive {MIMO} downlink
  with joint statistical beamforming and user scheduling},''
  \emph{\BIBforeignlanguage{en}{IEEE Trans. Wireless Commun.}}, vol.~16, no.~4,
  pp. 2181--2194, Apr. 2017.

\bibitem{christensen_weighted_2008}
S.~S. Christensen, R.~Agarwal, E.~De~Carvalho, and J.~M. Cioffi,
  ``\BIBforeignlanguage{en}{Weighted sum-rate maximization using weighted
  {MMSE} for {MIMO}-{BC} beamforming design},''
  \emph{\BIBforeignlanguage{en}{IEEE Trans. Wireless Commun.}}, vol.~7, no.~12,
  pp. 4792--4799, Dec. 2008.

\bibitem{smoothexp1}
Y.~Li, ``Simplified channel estimation for {OFDM} systems with multiple
  transmit antennas,'' \emph{IEEE Trans. Wireless Commun}, vol.~1, no.~1, p.
  67–75, 2002.

\bibitem{hu_wideband_2021}
X.~Hu and X.~Dai, ``\BIBforeignlanguage{en}{Wideband precoding for
  {MIMO}-{OFDM} systems with per-antenna power constraints},''
  \emph{\BIBforeignlanguage{en}{IEEE Commun. Lett.}}, vol.~25, no.~10, pp.
  3423--3426, Oct. 2021.

\bibitem{venugopal_optimal_2019}
K.~Venugopal, N.~Gonzalez-Prelcic, and R.~W. Heath,
  ``\BIBforeignlanguage{en}{Optimal frequency-flat precoding for
  frequency-selective millimeter wave channels},''
  \emph{\BIBforeignlanguage{en}{IEEE Trans. Wireless Commun.}}, vol.~18,
  no.~11, pp. 5098--5112, Nov. 2019.

\bibitem{9348441}
E.~Jeon, M.~Ahn, S.~Kim, W.~B. Lee, and J.~Kim, ``Joint beamformer and
  beamformee design for channel smoothing in {WLAN} systems,'' \emph{in Proc.
  IEEE 98th Veh. Technol. Conf.}, pp. 1--6, 2020.

\bibitem{smooth}
F.~L. W.~Hu and Y.~Jiang, ``Phase rotations of {SVD}-based precoders in
  {MIMO-OFDM} for improved channel estimation,'' \emph{IEEE Wireless Commun.
  Lett}, vol.~10, no.~8, p. 1805–1809, 2021.

\bibitem{9348}
Y.~Zhang, A.-A. Lu, {X. Q. Gao}, and X.-G. Xia, ``Cross-subcarrier precoder
  design for massive {MIMO-OFDM} downlink with symplectic optimization,''
  \emph{Sci. China Inf. Sci. under review}, 2024.

\bibitem{symplectic2}
M.~Betancourt, M.~I. Jordan, and A.~C. Wilson, ``On symplectic optimization,''
  \emph{arXiv:1802.03653v2}, 2018.

\bibitem{jordandan}
M.~Ghirardelli, ``Optimization via conformal hamiltonian systems on
  manifolds,'' \emph{Journal of Computational Dynamics}, vol.~11, no.~3, pp.
  354--375.

\bibitem{symplecticv1}
M.~I. Jordan, ``Dynamical, symplectic and stochastic perspectives on
  gradient-based optimization,'' \emph{Proc. Int. Cong. Of Math (ICM)}, pp.
  523--549, 2018.

\bibitem{symplectic3}
L.~C. Hairer, E. and G.~Wanner, ``Geometric numerical integration:
  Structure-preserving algorithms for ordinary differential equations,''
  \emph{Springer, New York}, 2006.

\bibitem{step}
N.~S. Wadia, M.~I. Jordan, and M.~Muehlebach, ``Optimization with adaptive step
  size selection from a dynamical systems perspective,'' \emph{Proceedings of
  the 35th Conference on Neural Information Processing Systems, NeurIPS
  Workshop on Optimization for Machine Learning}, 2021.

\bibitem{steprange}
E.~Hairer and G.~Wanner, ``Solving ordinary differential equations {II},''
  \emph{Springer Ser. Comput. Math., Springer-Verlag, Berlin}, 1996.

\bibitem{jaeckel_quadriga_2014}
S.~Jaeckel, L.~Raschkowski, K.~Borner, and L.~Thiele,
  ``\BIBforeignlanguage{en}{{Quadriga}: {A} 3-{D} multi-cell channel model with
  time evolution for enabling virtual field trials},''
  \emph{\BIBforeignlanguage{en}{IEEE Trans. Antennas Propag.}}, vol.~62, no.~6,
  pp. 3242--3256, Jun. 2014.

\bibitem{ANDERSEN198324}
H.~C. Andersen, ``Rattle: A “velocity” version of the shake algorithm for
  molecular dynamics calculations,'' \emph{Journal of Computational Physics},
  vol.~52, no.~1, pp. 24--34, 1983.

\end{thebibliography}

\end{document}